\documentclass[fleqn,twoside,twocolumn,nofootinbib,showkeys]{revtex4} % Specifies the document class %,unsortedaddress
\usepackage[nocpr]{ujp} % \usepackage[cyr]{ujp} for cyrillic, \usepackage[web]{ujp} for web
\begin{document}

\title[Dimerization of Water Molecules]
{DIMERIZATION OF WATER MOLECULES.\\ MODELING OF THE ATTRACTIVE PART\\
OF THE INTERPARTICLE POTENTIAL\\ IN THE MULTIPOLE APPROXIMATION}%

\author{P.V. Makhlaichuk, M.P. Malomuzh, I.V.~Zhyganiuk}%1
\affiliation{I.I. Mechnikov Odesa National  University, Dept. of Theoretical Physics}%1
\address{2, Dvoryans'ka Str., Odesa 65026, Ukraine}%1
\email{mahlaichuk@gmail.com}%e-mail 1

\udk{???}  \razd{\secix}

\autorcol{P.V.\hspace*{0.7mm}Makhlaichuk,
M.P.\hspace*{0.7mm}Malomuzh, I.V.\hspace*{0.7mm}Zhyganiuk}

\setcounter{page}{278}%

\begin{abstract}
The work presents the detailed analysis of the water dimer
properties. Their parameters are investigated on the basis of a
multipole interaction potential extended up to the
quadrupole--quadrupole and dipole--octupole terms. All main
equilibrium parameters of the dimer are obtained: its geometry,
ground-state energy, dipole and quadrupole moments, vibration
frequencies, {\it etc}. They are thoroughly compared with those obtained in
quantum chemical calculations and from spectroscopic data. The
efficiency of the present model potentials is discussed. A new
viewpoint on the nature of the hydrogen bond is presented. The results of
studies are thoroughly compared with the spectroscopic and computer
simulation data.
\end{abstract}

\maketitle

\section{Introduction}

The main object of our research is the water dimer properties, which
parameters are very sensitive to the type of intermolecular
interaction. A water dimer is a closed system of two water molecules
connected with the hydrogen bond. The equilibrium dimer
configuration that corresponds to the minimum of the interaction
energy is presented in Fig.~1. We will briefly discuss the main
dimer parameters presented in Table 1 according to the quantum
chemical calculations and experimental data. In \cite{SD}, two types
of dimers are studied: a ``frozen'' dimer, which has the distance
between oxygen and hydrogen and the angle between O--H bonds fixed,
and a ``relaxed'' dimer, where changes of the O--H length and the
angle between bonds are allowed. In \cite{Umeyama}, a ``frozen''
dimer configuration was used to determine the H-bond energy and the
dipole moment of a dimer (the intermolecular distance
$r_{{{\scriptscriptstyle{\mathrm O}} } {{\scriptscriptstyle{\mathrm
O}} } }^{(0)}$  and the angle $\theta_0$  were postulated). From the
comparison of the quantum chemical results and the experimental
ones, it follows that the optimum value of the ground-state energy
of a water dimer is $\Phi_d\approx-(9\div10)k_{\rm B}T_m$, where
$T_m=273$~K is the melting temperature of ice, and is observed at
$r_{{{\scriptscriptstyle{\mathrm O}} } {{\scriptscriptstyle{\mathrm
O}} } }^{(0)}=2.98$\,{\AA}. The values of angle $\theta_0$  and
dimer dipole moment $D_d$  should be $\theta_0=50^\circ$  and
$D_d\approx2.8$\,D, \mbox{respectively.}\looseness=1

%Tabl.1
\begin{table*}[!]
\vskip4mm \noindent\caption{ Characteristics of the equilibrium
state of a water dimer according  to different approaches
}\vskip3mm\tabcolsep15.7pt

\noindent{\footnotesize\begin{tabular}{|c|c|c|c|c|c|}
  \hline
  \parbox[c][9mm][c]{10mm}{} &
  \parbox[c][9mm][c]{40mm}{Specification } &
  \parbox[c][9mm][c]{17mm}{$r_{{{\scriptscriptstyle{\mathrm O}} } {{\scriptscriptstyle{\mathrm O}} } }^{(0)}$,\,\AA } &
  \parbox[c][9mm][c]{12mm}{$\theta_0$, deg } &
  \parbox[c][9mm][c]{17mm}{$E_d$} &
  \parbox[c][9mm][c]{10mm}{$D_d$, D} \\[2mm]
  \hline \rule{0pt}{5mm}\raisebox{-1.5mm}[0cm][0cm]{\cite{SD}}&Frozen dimer&2.954&19.3&--9.94&\\[-0.7mm] %
 &Relaxed dimer&2.896&20.5&--12.78&\\%[0.5mm]%
 &6-31G&2.98&60&--10.69&2.2\\[-0.7mm]%
\cite{Umeyama} &STO-3G&2.98&60&--9.39&1.72\\[-0.7mm]%
 &4-31G&2.98&60&--14.18&2.60\\%[0.5mm]%
 \cite{Schutz}& &2.925&51.8&--9.11&\\%[0.5mm]%
 \cite{Matsuoka}&&&--10.32&&\\%[0.5mm]%
 &HF G-311G&2.814& &--15.06&4.37\\[-0.7mm]%
&HF  G-311G++G(dp)  &3.001& &  --8.85&  3.45\\[-0.7mm]%
&HF  G-311G++G(3df, 3pd) &      3.026   &&  --7.37&  2.88\\[-0.7mm]%
\raisebox{-1.5mm}[0cm][0cm]{\cite{Lipkowitz}}&MP2  G-311G    &2.797  &&  --16.82  &4.24\\[-0.7mm]%
&MP2  G-311G++G(dp)&    2.914&&     --11.21& 3.30\\[-0.7mm]%
&MP2  G-311G++G(3df, 3pd)&  2.904&&     --9.79   &2.68\\[-0.7mm]%
&BLYP G-311G++G(dp)&    2.927&&     --10.03& 3.09\\[-0.7mm]%
&BLYP G-311G++G(3df, 3pd)&  2.944&&     --8.22&  2.54\\%[0.5mm]%
\cite{Yu}&&&&&2.6\\%[0.5mm]%
\cite{Silvestrelli}&&2.94&&&2.15\\%[0.5mm]%
\cite{Odutola}&Exper.&$2.976\!\pm\!0.004$&$57\!\pm\!10$&&\\%[0.5mm]%
\cite{Kessler}&Exper.&&&$-9.96\!\pm\!0.4$&\\%[0.5mm]%
\cite{Dyke}&Exper.&&&&2.60\\[2mm]
  \hline
\end{tabular}
  \label{gerpes}}
\end{table*}

This brief review of the dimer properties shows that, nowadays, there
is no consistent approach to the problem of formation of dimers, as
well as to the calculation of their equilibrium parameters. The
biggest difficulty is related to the absence of a clear approach
to the magnitude of characteristic contributions to the interaction
energy.

In the present work, the water dimer properties are studied on the basis of
a strictly defined potential that consists of: 1) literature-based
dispersive and repulsive interaction and 2) multipole electrostatic
interaction that is extrapolated to the overlapping region. Under the
sufficient proximity of water molecules that leads to the overlap
of the electron shells, it is necessary to use the quantum chemical
calculations to determine the interaction energy. It was shown in
\cite{Mahlaichuk} that the energy of a hydrogen bond itself does
not exceed $\sim k_{\rm B}T_m$. The last circumstance is in
agreement with the fact that the intramolecular distances in a water
molecule change not more than by $(1\div3)\%$
\cite{Odutola,Kessler}. This means that we can neglect the
contributions of hydrogen bonds to the interaction energy of two
water molecules. For the first time, this fact had been understood in
the works of Berendsen \cite{BerendsenPot} and Barnes \cite{Barnes}.
A critical review of the modern potentials, in which the existence of
hydrogen bonds is ignored, is presented in \cite{Umeyama}. In the
same work, the generalized Stillinger--David potential that
eliminates its weaknesses and retains its positive features is
presented. Let us mention the attempt
in \cite{Ichiyea} to model the hydrogen bonds, but no results that deserve a
significant attention are obtained. Specifically, the following aspects are
studied in the present work: 1)~the ground-state energy of a water
dimer along with its dipole moment as functions of
the intermolecular distance and the angles that define a relative
orientation of molecules; 2)~the influence of dimer's rotation on
its ground-state energy and 3)~vibration states of a water dimer. The
possibility to use the electrostatic multipole potential for the
description of attraction between water molecules in liquid water
and the relations between the results obtained, computer simulations,
and experimental data are discussed.

\section{Structure of the Interparticle Interaction Potential Between Water Molecules}

The interparticle interaction potential is modeled with the expression
%1
\begin{equation}
\Phi(r,\Omega)\!=\!\Phi_r(r,\Omega)\!+\!\Phi_D(r,\Omega)\!+\!\Phi_M(r,\Omega)\!+\!\Phi_{{\mathrm
H}}(r,\Omega),
\end{equation}
where $\Phi_r(r,\Omega)$  is the repulsive term, $\Phi_D(r,\Omega)$
describes the dispersive forces, $\Phi_M(r,\Omega)$ is a part of
the multipole expansion of the interaction energy between two water
molecules, and $\Phi_{{\mathrm H}}(r,\Omega)$ is the irreducible
contribution caused by the overlap of the electron shells of the water
molecules ($\Omega$ is the set of angles that describe the
orientation of water molecules). The reference to the multipole
expansion for the interparticle interaction is supported by the
following facts: 1) quantum chemical calculation of the multipole moments
is a well-posed problem; 2) comparison of the
different multipole contributions allows us to control the applicability
of the multipole approximation for the electrostatic interaction. Due
to the wide discussion of the parameters
of dimers \cite{Kistenmacher,Zhyganiuk,Schutz,Matsuoka,Odutola,Kessler,Makushkin,Lipkowitz,Yu,Silvestrelli},
the last fact becomes stronger.

In the present work, the multipole contributions to the interparticle
potential are considered up to the three-moment terms, i.e.
up to the octupole-octupole effects:
%2
\begin{equation}
\Phi_M(r)=\sum_{1\leq n,m \leq
3}\frac{(-1)^n}{n!m!}(\hat{Q}_1^{(n)}\hat{Q}_2^{(m)}):\hat{D}^{(n+m)}\frac{1}{r}.
\end{equation}

%Fig.1
\begin{figure}% figure* for wide figure, [h] [!] to change the placement
\vskip1mm
\includegraphics[width=3.8 cm]{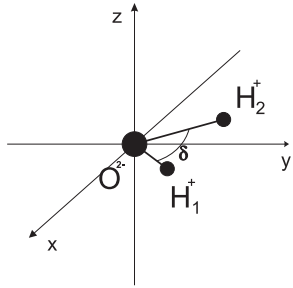}
\vskip-3mm\caption{ Molecular coordinate system }\vskip3mm
\end{figure}
%Fig.2
\begin{figure}% figure* for wide figure, [h] [!] to change the placement
\includegraphics[width=7.3 cm]{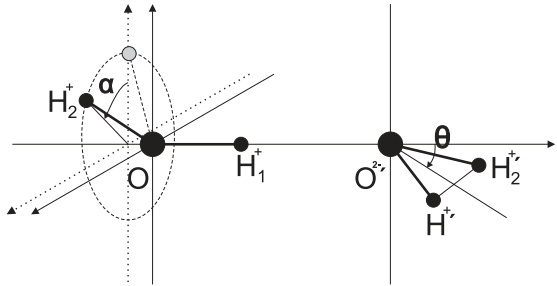}
\vskip-3mm\caption{ Angles that define the orientation of molecules
in LCS }
\end{figure}

%Tabl.2
\begin{table}[b]
\noindent\caption{ Components of the quadrupole\\ moment of a water
molecule in MCS }\vskip3mm\tabcolsep5.8pt

\noindent{\footnotesize\begin{tabular}{|c|c|c|c|}
  \hline
  \parbox[c][9mm][c]{30mm}{} &
  \parbox[c][9mm][c]{20mm}{$Q_{XX}^{(0)}$, D \AA } &
\parbox[c][9mm][c]{20mm}{$Q_{YY}^{(0)}$, D \AA } &
 \parbox[c][9mm][c]{20mm}{$Q_{ZZ}^{(0)}$, D \AA} \\[2mm]
  \hline \rule{0pt}{5mm}\cite{SD}&2.57&0.37&--2.94\\
\cite{Mahlaichuk}&2.34&0.43&--1.17\\%
\cite{Silvestrelli}&-0.09&--1.67&1.75\\%
\cite{Verhoeven}&--0.07&--1.61&1.69\\%
\cite{Aung}&--0.19&--1.58&1.77\\%
\cite{Harrison}&--0.07&--1.61&1.69\\%
\cite{Neumann}&--0.55&--0.84&1.39\\%
\cite{Cipriani}&--0.07&--1.74&1.81\\[2mm]
  \hline
\end{tabular}}
\end{table}

\noindent Here, $\hat{Q}^{(k)}$ is the $k$-th multipole moment,
whose components are defined as \mbox{$\hat{Q}^{(k)}\Rightarrow
$}\linebreak $\Rightarrow\sum_{1\leq i \leq N}q_i x^{(i)}_{\alpha_1}
x^{(i)}_{\alpha_2} ... x^{(i)}_{\alpha_k} $, $\hat{D}^{(k)}
\Rightarrow \frac{\partial^k}{\partial x_{\alpha_1} \partial
x_{\alpha_2} ...
\partial x_{\alpha_k} } $ is the differential operator and (:) stands for
the sum over the indexes. The symbol $(\hat{Q}_1^{(n)}
\hat{Q}_2^{(m)})$  is used to determine the Kronecker product of the
matrices that correspond to the multipole moments, and  $r$ is the
distance between oxygens. After all the necessary differentiations,
we substitute: $r \Rightarrow {\bf r}_{{{\scriptscriptstyle{\mathrm
O}} } {{\scriptscriptstyle{\mathrm O}} } }$ and $x_\alpha
\Rightarrow r_{{{\scriptscriptstyle{\mathrm O}} }
{{\scriptscriptstyle{\mathrm O}} } }\delta_{2\alpha},$ which
corresponds to a selected dimer configuration. Multipole
contributions include terms up to the dipole-octupole ones:
%3
\begin{equation}
\Phi_M(r)=\Phi_{DD}+\Phi_{DQ}+\Phi_{QQ}+\Phi_{\rm DO}+...,
\end{equation}
which take the form
\[\Phi_{DD}=\frac{1}{r_{{{\scriptscriptstyle{\mathrm O}} } {{\scriptscriptstyle{\mathrm O}} } }^3}
({\bf d}^{(1)}{\bf d}^{(2)}-3{\bf d}_2^{(1)}{\bf
d}_2^{(2)}),\]\vspace*{-5mm}
\[\Phi_{DQ}=-\frac{1}{2 r_{{{\scriptscriptstyle{\mathrm O}} } {{\scriptscriptstyle{\mathrm O}} } }^4}
 (6(d_{\alpha}^{(1)}Q_{2\alpha}^{(2)} + d_{\alpha}^{(2)}Q_{2\alpha}^{(1)})\,-\]\vspace*{-5mm}
\[-\,15(d_{2}^{(1)}Q_{22}^{(2)} + d_{2}^{(1)}Q_{22}^{(2)})),\]\vspace*{-5mm}
\[\Phi_{QQ}=\frac{3}{4r_{{{\scriptscriptstyle{\mathrm O}} } {{\scriptscriptstyle{\mathrm O}} } }^5}
(35 Q_{22}^{(1)}Q_{22}^{(2)} - 20 Q_{2\delta}^{(1)}Q_{2\delta}^{(2)}
+ 2 Q_{\delta \gamma}^{(1)}Q_{\delta \gamma}^{(2)}),\]\vspace*{-5mm}
\[\Phi_{\rm DO}=-\frac{3}{2r_{{{\scriptscriptstyle{\mathrm O}} }
{{\scriptscriptstyle{\mathrm O}} } }^5} ( d_\alpha^{(1)}O_{\alpha
\beta \beta}^{(2)} + d_\alpha^{(2)}O_{\alpha \beta \beta}^{(1)}
+\]\vspace*{-5mm}
\[+\, 15 ( d_2^{(1)}O_{222}^{(2)} + d_2^{(2)}O_{222}^{(1)} ) - 5 (
d_2^{(1)}O_{2\alpha \alpha}^{(2)} + d_2^{(1)}O_{2\alpha
\alpha}^{(1)})\,+
\]\vspace*{-5mm}
\[+\, d_\alpha^{(1)}O_{\alpha22}^{(2)}+d_\alpha^{(2)}O_{\alpha22}^{(1)} ).\]

The components of the dipole moment of a water molecule in the laboratory coordinate system (LCS) are given by the relations
\[{\bf d}^{(1)}=d(0,\cos(\delta/2+\chi),\sin(\delta/2+\chi)),\]\vspace*{-5mm}
\[{\bf d}^{(2)}=d(0,\cos\theta,-\sin\theta),\]
where $d$ is the absolute value of dipole moment of a water
molecule, $\delta$ is the angle in the molecular coordinate system
(MCS) that defines the positions of hydrogens. It is stated that
the water molecule is oriented in MCS as in Fig. 2.

The values of components of the quadrupole and octupole moments, calculated in computer experiments or
obtained experimentally are presented in Tables 2 and 3. The orders of magnitude for the multipole moments are:
\[d\sim 1~{\rm D},~~ Q\sim 10^{-8}~{\rm D}\cdot {\rm cm} ~~ {\rm and}~~ \ O\sim 10^{-16}~{\rm D}\cdot {\rm
cm}^2.
\]
Here, D equals 1 Debye -- unit of dipole moments. In our opinion,
the optimal values of components of the quadrupole moment of a water
molecule are obtained in the computer experiment
\cite{Silvestrelli}. These values are in good accordance with the
experimental data \cite{Verhoeven}. But the components of the
quadrupole moment that correspond to the charge distribution in
model potentials
\cite{SD,Berendsen1,Berendsen2,Jorgensen1,Jorgensen2} differ even by
the sign (see Table~2). The same situation is also characteristic of
the octupole moments (see Table~3). The mutual coherence of
different works is only observed in the values of dipole moment.
This fact gives us the ground to conclude that most of the model
potentials are unable to reproduce fine details of the
intermolecular interaction. The transition between MCS and LCS is
\mbox{standard}:\looseness=1
\[{\bf d}(\Omega)=R_\Omega{\bf d},\]\vspace*{-5mm}
\[Q_{\alpha \beta}(\Omega)=R_\Omega Q_{\alpha \beta}  R_\Omega^T,\]\vspace*{-5mm}
\[O_{Y\alpha \beta}(\Omega)=R_\Omega O_{Y\alpha \beta}  R_\Omega^T,\]

\noindent where $R_\Omega$ indicates the rotation matrix by the
angle $\Omega$: $\Omega=\alpha,\theta,\phi$ ($\alpha,\theta,\phi$
are the angles of rotation around the LCS axes, as given in Fig. 3)

For example, the rotation around the $Oy$ axis is given by the
direct and transposed matrices:
\begin{eqnarray*}
    R_\alpha\!=\! \begin{pmatrix} \cos \alpha & 0 & \sin \alpha \\ 0 & 1 & 0
    \\ -\sin \alpha & 0 & \cos \alpha \!\end{pmatrix}\!, ~~R_\alpha^T\!= \!\begin{pmatrix}
     \cos \alpha & 0 & -\sin \alpha \\ 0 & 1 & 0 \\ \sin \alpha & 0 & \cos \alpha
     \!\end{pmatrix}\!.
\end{eqnarray*}

We note that the components of the quadrupole moment of a water
molecule in [5, 15--20] were calculated with the use of the formula  $
\tilde{Q}_{\alpha \beta}^{(0)}=\frac{1}{2}\sum_i
(3x_\alpha^{(1)}x_\beta^{(i)}-$ $-{\bf r}_i^2 \delta_{\alpha \beta}
) $ that differs from our one by the multiplier 3/2.

According to the selection of LCS (see Fig. 1), we can use only
$O_{Y\alpha \beta}$ components of the octupole moment. By the same
reason, as we use traceless quadrupole moments, we will use traceless
octupole moments:\vspace*{-2mm}
\[O_{Y \alpha \beta}^{(0)}=O_{Y \alpha \beta}-\frac{1}{3}O_Y \delta_{\alpha \beta},\]
where  $O_Y=\sum_{\alpha=1}^3O_{Y \alpha \alpha} $ is the trace of
the matrix $O_{Y \alpha \beta}$. From our point of view, the most
convenient values of $O_{Y \alpha \beta}$ are obtained in
\cite{Silvestrelli} and \cite{Verhoeven}, where the values of
components of the octupole moments are almost the same.

%Fig.3
\begin{figure}% figure* for wide figure, [h] [!] to change the placement
\vskip1mm
\includegraphics[width=\column]{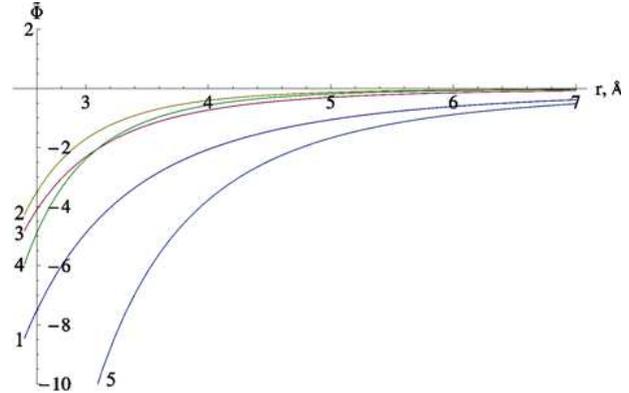}
\vskip-3mm\caption{ Role of the corresponding contributions to the
interparticle potential: {\it 1} -- dipole-dipole interaction, {\it
2} -- dipole-quadrupole, {\it 3} -- quadrupole-quadrupole, {\it 4}
-- dipole-octupole, {\it 5}~-- total multipole interaction
}\vskip3mm
\end{figure}
%Fig.4
\begin{figure}% figure* for wide figure, [h] [!] to change the placement
\includegraphics[width=\column]{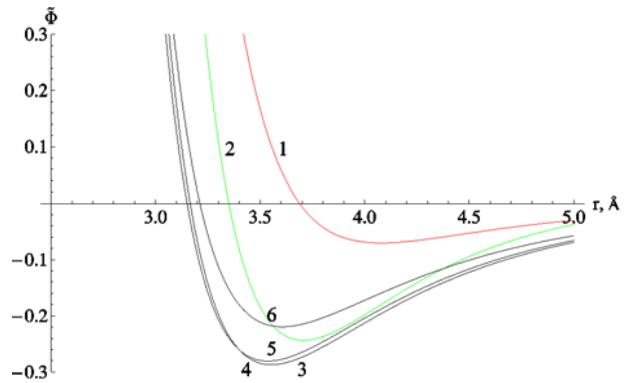}
\vskip-3mm\caption{ The behavior of the sum of repulsive and
dispersive terms for the Buckingham ({\it 1}), Buckingham--Corner
({\it 2}), SPC~({\it 3}), SPC/E ({\it 4}), TIPS ({\it 5}), and TIP3P
({\it 6}) potentials}
\end{figure}

%Tabl.3
\begin{table}[b]
\noindent\caption{ Components of the octupole\\ moment of a water
molecule }\vskip3mm\tabcolsep5.7pt

\noindent{\footnotesize\begin{tabular}{|c|c|c|c|}
  \hline
  \parbox[c][9mm][c]{30mm}{} &
  \parbox[c][9mm][c]{20mm}{$O_{YXX}^{(0)}$, D~{\AA}$^2$ } &
\parbox[c][9mm][c]{20mm}{$O_{YYY}^{(0)}$, D~{\AA}$^2$} &
\parbox[c][9mm][c]{20mm}{$O_{YXX}^{(0)}$, D~{\AA}$^2$} \\[2mm]
  \hline \rule{0pt}{5mm}\cite{SD}&1.91&0.30&--2.22\\%
\cite{Silvestrelli}&--1.29&--1.73&3.03\\%
\cite{Verhoeven}&--1.33&--1.82&3.16 \\[2mm]
  \hline
\end{tabular}
  \label{gerpes}}
\end{table}

The relative values of the multipole contributions of different orders
to the interparticle potential are presented in Fig. 4. Here and
below, we use the the dimensionless values for interaction energy
$\Phi(r,\Omega)\rightarrow $ $\rightarrow\tilde{\Phi}(r,\Omega) /
k_{\rm B}T_m $, where $T_m$ is the crystallization temperature for
liquid water. The curves in Fig. 4 correspond to the fixed
value of angle $\chi,$ which is supposed to equal $0.7^\circ$ and
does not depend on the distance between water molecules. The angle
$\theta$ is a function of the intermolecular distance and was
obtained from the minimum of the interaction energy at each value of
$r_{\rm OO}$. The total value of multipole contribution is
shown by curve {\it 5}. As we see, at a distance of
3~{\AA} that is supposed to be the equilibrium for a water dimer,
the depth of the total multipole interaction reaches $-15$.

\begin{figure}% figure* for wide figure, [h] [!] to change the placement
\vskip1mm
\includegraphics[width=\column]{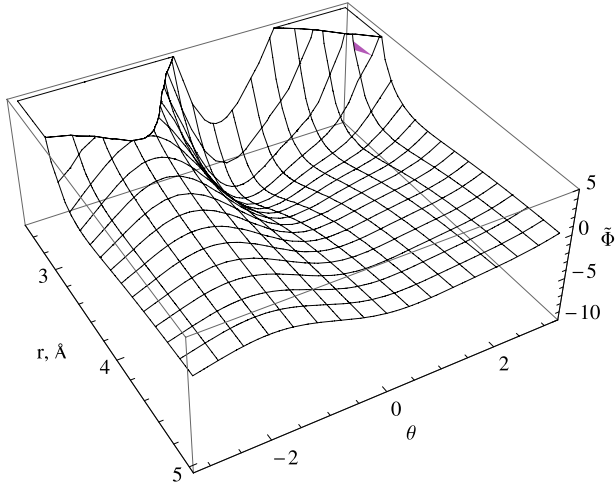}
\vskip-3mm\caption{Profile of the potential surface
$\tilde{\Phi}=\Phi/k_{\rm B}T_m$ of the interaction of two water
molecules ( $T_m$ is the temperature of melting for ice)}
\end{figure}

We should consider the behavior of the  $\Phi_{\rm DO}$ term that
describes the dipole-octupole interaction. From Fig. 4, it follows
that the curves of $\Phi_{\rm DO}$  and $\Phi_{QQ}$ cross at $r\sim
r_\ast$, where $r_\ast \approx 3.2$~{\AA}. This fact indicates the
breaking of the multipole expansion. At the smaller distances, the
electron shells begin to overlap, so it is the natural border for
the usage of model electrostatic potentials.

\section{Repulsive and Dispersive Interaction between Water Molecules}

We will use the Buckingham--Corner \cite{Eizenberg}, three-point
SPC \cite{Berendsen1,Berendsen2}, and TIPS
\cite{Jorgensen1,Jorgensen2} potentials to describe the repulsion
and dispersive effects between water molecules. In the
Buckingham--Corner potential, the effects of repulsion and dispersive
forces are described with the sum of pair contributions of
the hydrogen--hydrogen, hydrogen--oxygen, and oxygen--oxygen types:
%4
\begin{equation}
\tilde{\Phi}(r)=\tilde{B}e^{-\tilde{p}(\frac{r}{r_m})} -
\left(\!\frac{\tilde{A}_6}{r^6}+\frac{\tilde{A}_8}{r^8}\!\right)e^{-4(\frac{r}{r_m}-1)^3}\!
,
\end{equation}
where
$\tilde{B}=(-\epsilon+(1+\beta)\frac{\tilde{A}_6}{r_m^6})e^{-\tilde{p}}$,
$\tilde{A}_6=\frac{\tilde{\epsilon}\tilde{p}r_m^6}{\tilde{p}(1+\beta)-8\beta-6}
$, $\tilde{A}_8=\beta r_m^2 \tilde{A}_6 $,
$\beta=\frac{\tilde{A}_8r_m^{-8}}{\tilde{A}_6r_m^{-6}}$,
$\epsilon$ is the potential well depth, and $r_m$ is the
intermolecular distance at the minima of energy.

In the SPC and TIPS potentials, the dispersive and repulsive
interactions inhere only in the electron shells of the oxygen
atoms and are determined in the form similar to the Lennard-Jones
potential:
%5
\begin{equation}
\Phi_r^{\rm (SPC)}=\frac{\tilde{B}_{{{\scriptscriptstyle{\mathrm O}}
} {{\scriptscriptstyle{\mathrm O}} }
}}{r_{{{\scriptscriptstyle{\mathrm O}} }
{{\scriptscriptstyle{\mathrm O}} } }^{12}}, \quad \Phi_d^{\rm
(SPC)}=-\frac{\tilde{A}_{{{\scriptscriptstyle{\mathrm O}} }
{{\scriptscriptstyle{\mathrm O}} }
}}{r_{{{\scriptscriptstyle{\mathrm O}} }
{{\scriptscriptstyle{\mathrm O}} } }^6}.
\end{equation}

The comparative behavior of the repulsive and dispersive
contributions to the Buckingham, Buc\-king\-ham--Corner, SPC, TIPS,
and TIP3P potentials is presented in Fig. 4.

As we can see, the minimum of the interaction energy is around
$0.2\div0.3,$ and its position is approximately at
$(3.5\div3.6)$~{\AA}. Curve {\it 1} that corresponds to the
Buckingham potential differs very much from the latter potentials.
Of course, the simple comparison of the dispersive and repulsive
curves is not enough to select the appropriate potential.

\section{Ground State of a Dimer }

In this section, we will present the results of studies of the
ground state of a water dimer, based on potential (1), where we
will ignore the contribution of the short-range H-bond potential.
We will consider the positions of the oxygens and hydrogens in water
molecule remain still for the ease of calculations. According to
\cite{Zhyganiuk}, such requirement leads to the error not more than
$(1.5\div3)\%$. The ground state of a water dimer is identified with
the minimum of the interaction energy of two water molecules
oriented, as it is shown in Fig. 1. More specifically, the
equilibrium distance $r_{{{\scriptscriptstyle{\mathrm O}} }
{{\scriptscriptstyle{\mathrm O}} } }^{(0)}$  between the oxygens and
the angle $\theta_0$ between the directions of unexcited dipole
moments are found from the condition
\[r_{{{\scriptscriptstyle{\mathrm O}} } {{\scriptscriptstyle{\mathrm O}} } }^{(0)},
 \theta_0 ~~\leftrightarrow~~\text{absolute min}~~\Phi(\tilde{r},\theta,\alpha =
 0),
\]
where $\Phi$ is the intermolecular interaction potential. It is
considered to be a function of the dimensionless distance
$\tilde{r}=r_{{{\scriptscriptstyle{\mathrm O}} }
{{\scriptscriptstyle{\mathrm O}} } }/r_{{{\scriptscriptstyle{\mathrm
O}} } {{\scriptscriptstyle{\mathrm H}} } } $,
$r_{{{\scriptscriptstyle{\mathrm O}} } {{\scriptscriptstyle{\mathrm
H}} } }=0.97$~{\AA} is the length of the O--H bond,
and the angle $\alpha$ describes the rotation around the H-bond (see Fig. 2).

The general view of the potential surface  $\Phi(\tilde{r},\theta)$
is presented in Fig. 5.

On the set of points of minimum for $\Phi(\tilde{r},\theta),$
the angle $\theta_m$ is a function of $\tilde{r}$. Its dependence is
presented in Fig. 6.

The radial dependence of the interaction energy at the angle $\theta_m$
corresponding to the absolute minimum is presented in Fig. 7.

It follows from Figs. 5--7 that the parameters presented in Table 4
correspond to the absolute minimum of the interaction energy.

One can see that the equilibrium parameters of a dimer are very
sensitive to the selection of the form of repulsive and dispersive
interactions. More precisely, these parameters depend on the law of
decrease of the repulsive interaction. The depth of the potential
well becomes bigger when the maximum of repulsion moves to the left
(see Fig. 4). The position of the repulsive branch of the potential
is a key factor that defines the equilibrium parameters of a water
dimer. The value of the dipole moment of a water dimer is given by
the formula
%6
\begin{equation}\label{Dd}
D_d=2d_w \cos \frac{1}{2}(\theta_0+(\delta/2-\chi)).
\end{equation}

The values given should be compared to the equilibrium values
obtained from the computer simulations and experiments (see Table
1). As follows from Table 1, the intermolecular distance deviates
from the literature data. Another important dependence is the
dependence on the angle $\alpha$ of rotation around the H-bond
(see Fig.\,\,9). The variable $\alpha$  is an intradimer
characteristic, so it is responsive for the rotation of molecules
inside the dimer.

Another important characteristic of a dimer, which should be compared
with other results \cite{Makushkin,Lipkowitz,Yu}, is the dependence
on the angle $\theta$ corresponding to the transversal
vibrations of H-bonds (see Chapter 4). It is presented in Fig. 8. We
see that, in the region of the minimum, the dependence of the interaction
potential has a slope. As a consequence, in different models of water,
the deviations of the equilibrium value of $\theta$ can be huge.
Moreover, the heat vibrations of the H-bond can influence the
experimental values of the angle $\theta$.

%Fig.6
\begin{figure}% figure* for wide figure, [h] [!] to change the placement
\vskip1mm
\includegraphics[width=\column]{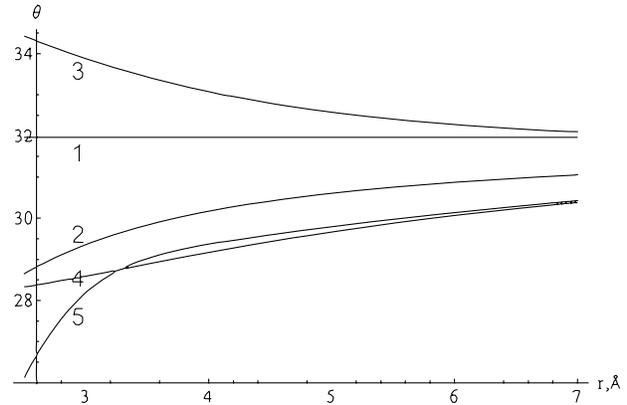}
\vskip-3mm\caption{Radial dependence of the angle $\theta_m$ ({\it
5}). Angles $\alpha$, $\beta,$ and $\chi$  remain fixed and
correspond to the absolute minimum configuration. Curves ${\it 1-4}$
describe the dependence of $\theta_m$ in $DD$, $DQ$, $QQ,$ and $DO$
approximations, respectively}\vskip3mm
\end{figure}
%Fig.7
\begin{figure}% figure* for wide figure, [h] [!] to change the placement
\includegraphics[width=\column]{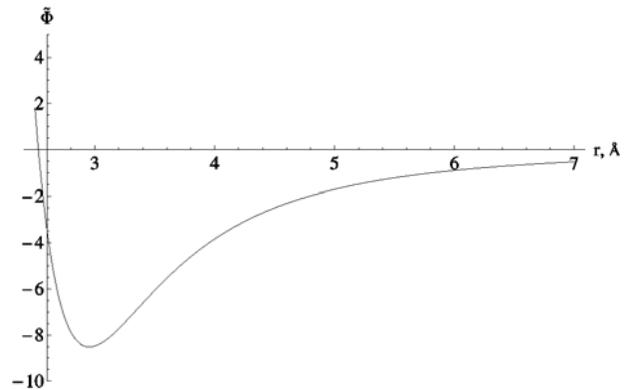}
\vskip-3mm\caption{Dependence of the interaction energy on the
intermolecular distance in a dimer configuration}
\end{figure}

%Tabl.4
\begin{table}[h!]
\vskip7mm \noindent\caption{ Equilibrium parameters of a water\\
dimer for different interaction potentials }\vskip3mm\tabcolsep5.0pt

\noindent{\footnotesize\begin{tabular}{|c|c|c|c|c|c|}
  \hline
  \parbox[c][9mm][c]{10mm}{} &
  \parbox[c][9mm][c]{7mm}{$r_{\rm OO}$} &
\parbox[c][9mm][c]{13mm}{$E_d,k_{\rm B}T_m$ } &
\parbox[c][9mm][c]{10mm}{$\theta$ } &
\parbox[c][9mm][c]{7mm}{$\chi$ } &
\parbox[c][9mm][c]{13mm}{$D_d$, D} \\[2mm]
  \hline \rule{0pt}{5mm}B&2.96&--8.5&28.03&0.75&2.8435\\%
BC&2.89&--9.65&28.70&0&2.8148\\%
SPC&2.72&--12.49&27.69&1.03&2.8143\\%
SPC/E&2.72&--12.49&27.69&1.03&2.8143\\%
TIPS&2.7&--12.66&27.61&1.14&2.8137\\%
TIP3P&2.69&--12.90&27.60&1.16&2.8136\\[2mm]
  \hline
\end{tabular}}
\end{table}

It is necessary to mention that the most convenient dimer parameters
to compare are: 1) the value of its dipole moment; 2) the values of
components of the quadrupole moment and it average value, which is
defined as $\bar{Q}=\frac{1}{3}(Q_{XX}+Q_{YY}+Q_{ZZ}) $ (here, we
use the components of the non-traceless quadrupole moment), and 3)
the vibration frequencies of water dimer. According to (\ref{Dd}),
the experimental value of the dimer dipole moment allows us to
control the equilibrium values of the angle $\theta$. The components
of the quadrupole moment of a water dimer are connected %
%Fig.8
\begin{figure}% figure* for wide figure, [h] [!] to change the placement
\vskip1mm
\includegraphics[width=\column]{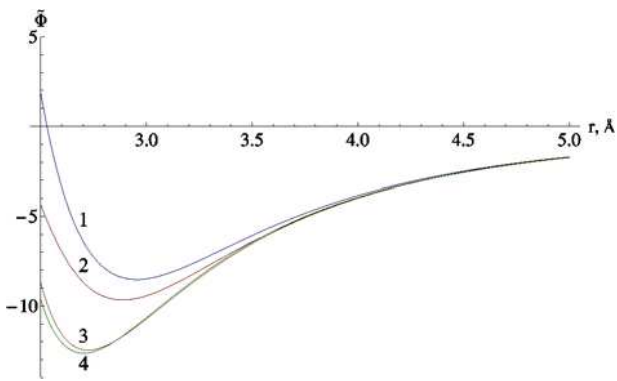}
\vskip-3mm\caption{Comparative behavior of the interaction
potentials. ({\it 1}) -- B, ({\it 2}) -- BC, ({\it 3}) -- SPC, ({\it
4}) -- TIPS}\vskip3mm
\end{figure}
%Fig.9
\begin{figure}% figure* for wide figure, [h] [!] to change the placement
\includegraphics[width=\column]{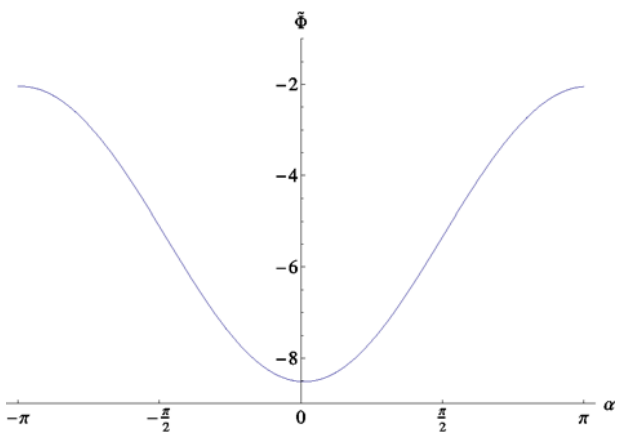}
\vskip-3mm\caption{Dependence of the dimer energy on the angle of
relative rotation around the H-bond}\vskip5mm
\end{figure}
%Fig.10
\begin{figure}% figure* for wide figure, [h] [!] to change the placement
\vskip1mm
\includegraphics[width=\column]{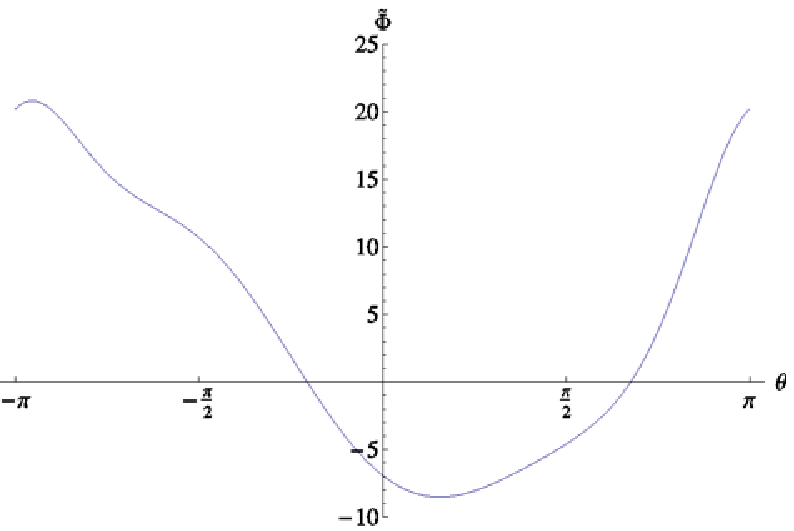}
\vskip-3mm\caption{Dependence of the interaction energy on the angle
$\theta$ at $r_{{{\scriptscriptstyle{\mathrm O}} }
{{\scriptscriptstyle{\mathrm O}} } }=2.9757$~{\AA}}\vskip3mm
\end{figure}
%Fig.11
\begin{figure}% figure* for wide figure, [h] [!] to change the placement
\vskip1mm
\includegraphics[width=\column]{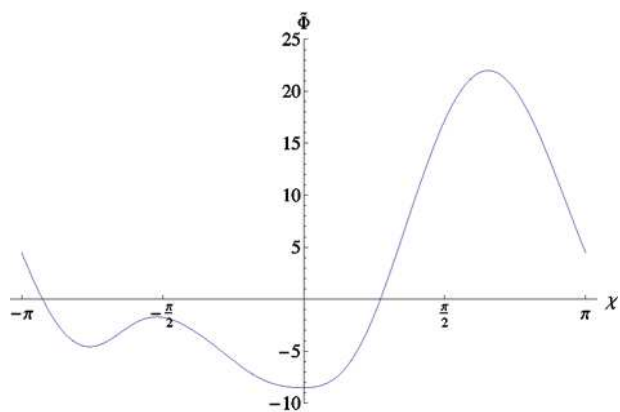}
\vskip-3mm\caption{Dependence of the interaction energy on the angle
$\chi$ }\vspace*{3mm}
\end{figure}
%Fig.12
\begin{figure}% figure* for wide figure, [h] [!] to change the placement
\includegraphics[width=\column]{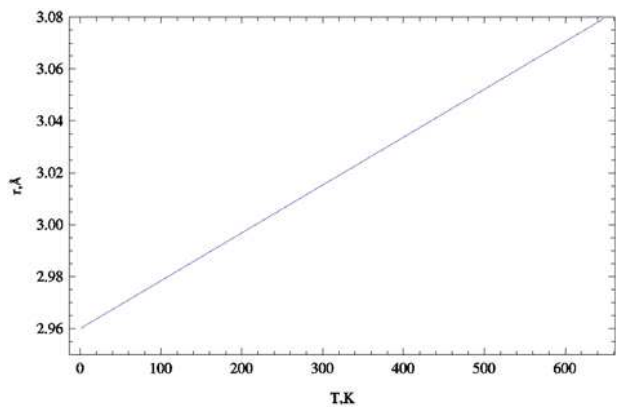}
\vskip-3mm\caption{Temperature dependence of the equilibrium
distance between oxygen atoms in a water dimer}\vspace*{3mm}
\end{figure}
%Fig.13
\begin{figure}[h!]% figure* for wide figure, [h] [!] to change the placement
\includegraphics[width=4.5 cm]{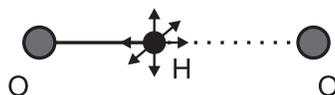}
\vskip-3mm\caption{Vibration modes of an H-bond}
\end{figure}%
with the components of the quadrupole moment of an isolated water
molecule by the relation
%7
\[
Q_{\alpha \beta}^{(d)}=Q_{\alpha \beta}^{(1)}+Q_{\alpha \beta}^{(2)}
+\frac{1}{2}(R_\alpha({\bf d}_2^{(w)}-{\bf d}_1^{(w)})_\beta\, +
\]\vspace*{-5mm}
\begin{equation}
+\,R_\beta({\bf d}_2^{(w)}-{\bf d}_1^{(w)})_\alpha),
\end{equation}

\noindent where $R=r_{{{\scriptscriptstyle{\mathrm O}} }
{{\scriptscriptstyle{\mathrm O}} } }^{(0)}$. The average value of
quadrupole moment is
%8
\begin{equation}
\bar{Q}_d=\bar{Q}_1+\bar{Q}_2+\frac{1}{3} d_w
r_{{{\scriptscriptstyle{\mathrm O}} } {{\scriptscriptstyle{\mathrm
O}} } }^{(0)}(\cos \theta - \cos \delta/2).
\end{equation}

%Tabl.5
\begin{table*}[!]
\vskip4mm \noindent\caption{Frequencies of small oscillations
}\vskip3mm\tabcolsep11.6pt

\noindent{\footnotesize\begin{tabular}{|c|c|c|c|c|c|c|}
  \hline
 &
 \multicolumn{2}{|c}{\rule{0pt}{5mm}B}&\multicolumn{2}{|c}{BC}&\multicolumn{1}{|c}{\raisebox{-4mm}[0cm][0cm]{SPC}}
 &\multicolumn{1}{|c|}{\raisebox{-4mm}[0cm][0cm]{TIPS}}\\[2mm]
 \cline{2-5}
  \parbox[c][9mm][c]{30mm}{} &
  \parbox[c][9mm][c]{20mm}{``Frozen dimer''} &
  \parbox[c][9mm][c]{20mm}{Rotating dimer } &
  \parbox[c][9mm][c]{20mm}{``Frozen dimer''} &
  \parbox[c][9mm][c]{20mm}{Rotating dimer } &&\\[2mm]
  \hline \rule{0pt}{5mm}$\omega_1,~{\rm cm}^{-1}$&175.12&166.23&182.51&164.20&63.14&68.91\\%
$\omega_2,~{\rm cm}^{-1}$&196.82&194.44&198.14&194.43&244.63&212.63\\%
$\omega_3,~{\rm cm}^{-1}$&391.83&385.63&400.87&386.73&246.00&247.64\\%
$\omega_4,~{\rm cm}^{-1}$&521.64&513.29&529.97&511.42&307.93&313.00\\[2mm]
  \hline
\end{tabular}}\vspace*{-2mm}
\end{table*}

It is worth to mention that the components of the dipole and quadrupole moments
can be easily calculated with the help of the methods of quantum
chemistry. Another surprising fact is the possibility of the
existence of metastable states of water dimers. With the fixed angle
$\theta$ that corresponds to a minimum, there is a possibility of
the turn of another water molecule on a specific angle, that can be
calculated studying the dependence of the interaction energy \mbox{on
angle $\chi$.}%\looseness=1

As follows from Fig. 11, there are two local minima separated
by the potential barrier. The depth of the second minimum is
approximately  $5k_{\rm B}T_m,$ and its position corresponds to the
reflection of a molecule relative to the \textit{xy} plane.

\section{Rotation of a Dimer}

If a dimer is formed in the gas phase, its thermal motion consists
of the translational, rotational, and vibration modes. It is clear
that only rotations and vibrations influence the equilibrium
parameters of a dimer. Let us mention that the experimental values
of intermolecular distance \cite{Odutola,Kessler,Dyke} correspond
to the rotating dimers. This circumstance is ignored in the
computer experiments. In this section, we will consider only the
rotational influence on the parameters of a dimer. The rotation of
a dimer is followed by the emergence of centrifugal forces, which
lead to the growth of the equilibrium distance between oxygens.
Rotation takes place around two axes that are perpendicular to the
$OX$ axis. Both axes lie in the molecular planes. The growth of
the distance between oxygens is determined by the equation
%9
\begin{equation}
    k(r-r_0)=m_0(\omega_1^2+\omega_2^2)r,
\end{equation}
where the average frequencies of rotation can be estimated from the relations
\[ \omega_1^2=\frac{k_{\rm B}T}{I_1}, \quad\omega_2^2=\frac{k_{\rm
   B}T}{I_2}.\]
The inertia moments  $I_1$ and $I_2$ are equal to: $I_1=$
$=1.332\times 10^{-33}~ {\rm g}\cdot {\rm cm}^2$, $I_2=1.219\times
10^{-33}~ {\rm g}\cdot {\rm cm}^2$. The values of force constant
\textit{k} are calculated in the next chapter. But, regardless the
calculations, the equilibrium distance grows linearly with the
temperature. The rise of the equilibrium distance between oxygens is
presented in Fig.~12.

As we can see, the distance between oxygens at the melting
temperature increases approximately by 0.04~{\AA}: from 2.96~{\AA}
to 3.00~{\AA}. The ground-state energy decreases by 0.04. In a
vicinity of the critical point, the equilibrium distance reaches
3.06~{\AA}.\vspace*{-2mm}

\section{Small Oscillations of a Dimer}
From the configuration of a dimer, it follows that the most
characteristic types of small oscillations are: 1)~longitudinal one
connected with a change of the  $r_{{{\scriptscriptstyle{\mathrm
O}} } {{\scriptscriptstyle{\mathrm O}} } }$ distance between the
oxygens: $\tilde{r}=\frac{r_{{{\scriptscriptstyle{\mathrm O}} }
{{\scriptscriptstyle{\mathrm O}} } }-r_{{{\scriptscriptstyle{\mathrm
O}} } {{\scriptscriptstyle{\mathrm O}} }
}^{(0)}}{r_{{{\scriptscriptstyle{\mathrm O}} }
{{\scriptscriptstyle{\mathrm H}} } }}$; 2)~two transversal
oscillations of the H-bond that are connected with small
rotations around the $x$ (by angle $\theta$) and $z$ (by angle
$\phi$ ) axes and 3)~intradimer oscillations connected with a
change of the relative orientation of water molecules during the
rotation around H-bonds. Here, it is necessary to mention that the
variables $\tilde{r}$ and $\theta$  are not independent and form a
new pair of normal (hybrid) coordinates. Longitudinal and two
transversal vibrations can also be interpreted as the oscillations
of the H-bond (Fig. 13). We have to remember that dimers rotate;
therefore, the spectra of small oscillations will change.

According to this, the Lagrange function for small oscillations of the dimer takes the form
%10
\begin{equation}
L=\frac{1}{2}\sum_{1\leq i,j \leq 4}m_{i,j}\dot{x}_i \dot{x}_j -
\frac{1}{2}\sum_{1\leq i,j \leq 4}K_{i,j}x_i x_j,
\end{equation}
where $x_i=(\tilde{r},\theta,\phi,\alpha)$ are the generalized
coordinates, and
$\dot{x}_i=(\dot{\tilde{r}},\dot{\theta},\dot{\phi},\dot{\alpha}) $
are the generalized velocities. It is easy to see that the mass
tensor has a diagonal structure:
\[ m_{11}=I_r, ~~m_{22}=I_\theta, ~~m_{33}=I_\phi,
~~m_{44}=I_\alpha,
\]
where $I_r=\mu r_{{{\scriptscriptstyle{\mathrm O}} }
{{\scriptscriptstyle{\mathrm H}} } }^2$, $\mu = 1/2 m_W$ is the
reduced mass of a water molecule,
$\tilde{I}=\frac{I_i^{(1)}I_i^{(2)}}{I_i^{(1)}+I_i^{(2)}}$  are the
components of the reduced inertia moments, $i=\theta,\alpha,$  and
the superscripts number water molecules. In accordance with Fig. 1,
the components of the inertia \mbox{moment are:}
\[I_{\theta}^{(1)}=2 m_{{\scriptscriptstyle{\mathrm H}} }
r_{{{\scriptscriptstyle{\mathrm O}} } {{\scriptscriptstyle{\mathrm
H}} } }^2;\]\vspace*{-9mm}
\[I_{\theta}^{(2)}=2m_{{\scriptscriptstyle{\mathrm H}} }
r_{{{\scriptscriptstyle{\mathrm O}} } {{\scriptscriptstyle{\mathrm
H}} } }^2\cos^2\delta/2;\]\vspace*{-9mm}
\[I_{\alpha}^{(1)}=m_{{\scriptscriptstyle{\mathrm H}} }
 r_{{{\scriptscriptstyle{\mathrm O}} } {{\scriptscriptstyle{\mathrm H}} } }^2(1+\cos^2\delta);\]\vspace*{-9mm}
\[I_{\alpha}^{(2)}=2m_{{\scriptscriptstyle{\mathrm H}} }
r_{{{\scriptscriptstyle{\mathrm O}} } {{\scriptscriptstyle{\mathrm
H}} } }^2(1-\cos^2\delta/2\cos^2\theta)\]
and are equal to
\[I_r\approx14.14\times10^{-46}~{\rm g}\cdot {\rm cm},\]\vspace*{-9mm}
\[\tilde{I}_\theta\approx0.83\times10^{-46}~{\rm g}\cdot {\rm cm},\]\vspace*{-9mm}
\[\tilde{I}_\phi\approx0.61\times10^{-46}~{\rm g}\cdot {\rm cm},\]\vspace*{-9mm}
\[\tilde{I}_\alpha\approx0.95\times10^{-46}~{\rm g}\cdot {\rm cm}.\]
The force constants  $k_{ij}$ are determined in a standard way:
$k_{ij}=\frac{\partial^2\tilde{\Phi}}{\partial x_i \partial
x_j}|_{x_i=0}$. All the derivatives are calculated at a fixed
value of angle $\chi$. The frequencies of small oscillations are
calculated in a standard way and are presented in Table 5 for the potential with repulsive
and dispersive parts from the Buckingham and Buckingham--Corner
potentials. Two first frequencies
correspond to hybrid oscillations of the $\tilde{r},\theta$-type.
Their normal coordinates are (at $\chi=0$):
%11
\begin{equation}
u_1=\frac{(k_{11}-I_r\omega_2^2)\tilde{r}+k_{12}\theta}{I_r(\omega_1^2-\omega_2^2)
(k_{22}-I_\theta\omega_1^2)}C_1\exp(i\omega_1t),
\end{equation}\vspace*{-5mm}
%12
\begin{equation}
u_2=\frac{k_{12}\tilde{r}+(k_{22}-I_\theta\omega_1^2)\theta}
{I_r(\omega_1^2-\omega_2^2)(k_{22}-I_\theta\omega_1^2)}C_2\exp(i\omega_2t).
\end{equation}
Coefficients  $C_1$ and $C_2$  are determined from the initial
conditions. Thus, we can  only speak about a single transversal
oscillation of an H-bond, while the other one mixes with its
longitudinal oscillation.%\vspace*{-2mm}

\section{Discussion}

From the qualitative point of view, the formation of dimers is
related to the formation of a hydrogen bond between two water
molecules. In this case, the ground-state energy of a dimer is
identified with the energy of the H-bond. In our dimensionless
units, it should be equal to --10. This estimate is confirmed by the
quantum chemical calculations \cite{Zhyganiuk,Schutz,Matsuoka}. The
close value of bonding energy ($\tilde{E}_d=-9.19$) is obtained, by
using a generalized Stillinger--David potential \cite{Zhyganiuk}.
However, the different approach is considered in the present work.
It is supposed that the dimers are formed due to three well-defined
types of interaction: repulsive, dispersive, and multipole
electrostatic interactions. In our calculation, we restrict
ourselves to all contributions up to the dipole-octupole ones. The
extrapolation of the interaction to the overlapping region does not
lead to any anomalous increase of the dipole-octupole interaction
relative to other terms. This is the evidence of the legitimacy of
using the described procedure of extrapolation on the distances up
to $r_{{{\scriptscriptstyle{\mathrm O}} }
{{\scriptscriptstyle{\mathrm O}} } }=3$~{\AA}. The reliability of
the results is confirmed by the fact that the values of quadrupole
and octupole moments calculated in the computer experiments coincide
with the experimental data. The optimal value of ground-state energy
of a stationary dimer according to our calculations is
$\tilde{E}_d=-9.65$ and is reached at the distance
$r_{{{\scriptscriptstyle{\mathrm O}} } {{\scriptscriptstyle{\mathrm
O}} } }^{(0)}=2.89$~{\AA}. It is shown that the rotation of a dimer,
which is quite natural in the gas phase, leads to an increase of the
equilibrium distance by $0.04$~{\AA}. One more circumstance that can
influence the experimental value of distance between oxygens is the
excitation of longitudinal oscillations that correspond to the
frequency $\omega_1$. By the order of magnitude, the amplitude of
oscillations \mbox{of dimers is}\looseness=1
\[ |\Delta r| \approx \sqrt{\frac{k_{\rm B}T_m}{m\omega_{||}^2}} \approx 0.1.\]%\vspace*{-4mm}

\noindent We see that the rotation and oscillations can change the
distance between oxygens by $0.1$~{\AA}. This fact cannot be
ignored. Another parameter of a dimer that can be controlled is the
angle $\theta_0,$ which defines the equilibrium mutual orientation
of the dipole moments of water molecules that form the dimer. This
angle is directly connected with the values of dipole and quadrupole
moments of a dimer, which can be calculated with the help of quantum
chemical methods or be obtained experimentally. In our work, the
following values of the angle and dipole moment of a dimer
correspond to the above interaction energy and the equilibrium
distance: $\theta_0=28.7^\circ$, $D_d=2.81$~D. These values slightly
differ from those obtained experimentally or calculated within the
quantum chemical methods (see Table~1). In
\cite{Makushkin,Lipkowitz}, it was shown that quantum chemical
calculations result in the dipole moment of a dimer $D_d=2.6$~D.
According to (\ref{Dd}), this value corresponds to the angle
$\theta_0\approx36^{\circ}$. The value of $\theta_0$  was also
studied in [7, 8]. It was shown that, at
$r_{{{\scriptscriptstyle{\mathrm O}} } {{\scriptscriptstyle{\mathrm
O}} } }=2.976$~{\AA}, the angle $\theta_0$ equals
$57^\circ\pm10^\circ$. From the spectroscopic experiments, it
follows that the angle should be $51^\circ\pm20^\circ$. Such a
difference in the values of $\theta_0$ is related to the sloping
character of the dependence $\tilde{\Phi}(\theta)$ in a vicinity of
the minimum, as it follows from Fig.~8. Vibrations and the rotation
of a dimer also influence the value of $\theta_0$:%\vspace*{-1mm}
\[ |\Delta\theta|\approx\sqrt{\frac{k_{\rm B}T_m}{I_\theta\omega^2}}\approx2.6^\circ .\]%\vspace*{-4mm}

\noindent The influence of the interaction of water molecules on the
parameters of water molecules will affect, in some way, the
equilibrium parameters of a dimer. It was demonstrated in [5] that
the dipole moment of a water molecule increases by $1\div 1.5\%$,
when molecules approach it to a distance of 3~{\AA}. So the growth
of the dipole-dipole interaction will result in changes of the
intermolecular distance and the interaction energy. The overlapping
effects that manifest themselves at the distances smaller than
3~{\AA} should not be considered because the relative shift of
valence vibrations of a proton at the condensation does not exceed
$1\div 3\%$ [14]. But the existence of hydrogen bonds should not be
neglected because they influence the value of heat capacity
\cite{PLA}. The weak overlapping of the electron shells shows itself
in the tunneling of protons from one molecule to the other one along
the hydrogen bond. But these effects are characteristic only at the
superlow frequencies ($\sim$30~${\rm cm}^{-1}$). All these
qualitative arguments allow us to state that the energy of the
hydrogen bond itself does not exceed $k_{\rm B}T_m$. The relatively
small contribution of hydrogen bonds to the interaction potential
was stated in the works by Dolgushin \cite{Dolgushin76,
Dolgushin77}. He showed that the sharing indices of the electron
density between water molecules under the influence of a neighbor
molecule are less than 3\%. These results were confirmed by the
later work by Fulton \cite{Fulton}. We mention that, in
\cite{Barnes, BerendsenPot}, hydrogen bonds were ignored without any
justification. In those works, it was shown that the interaction
energy calculated with the potential consisting only of the
classical electrostatic and repulsive interactions matches the
quantum chemical result with sufficient accuracy. The multipole
approximation allows us to easily build the averaged interaction
potential between water molecules. It should be noted that certain
limitations of the rigid multipole interaction will manifest
themselves in the description of multimer properties, where
multipartical effects take place. In particular, the equilibrium
distance between oxygens reduces to $r_{{{\scriptscriptstyle{\mathrm
O}} } {{\scriptscriptstyle{\mathrm O}} } }=2.78$~{\AA} in liquid
water. The denial of the model of hydrogen bond as the specific
donor-acceptor type of interaction finds its confirmation in another
fact. The values of self-diffusion coefficients and shear viscosity
for water have the same order of magnitude as those for the liquids
that do not form hydrogen bonds. This fact indicates that the
character of translational and rotational motions in these liquids
is similar. It cannot be agreed with the existence of sharply
directed hydrogen bonds with the interaction energy of $\sim$$
10k_{\rm B}T_m$. Our approach does not have such complexity. The
electrostatic multipole interaction satisfies the superposition
principle, so the total electric field formed by the randomly
positioned and orientated water molecules is relatively small and
does not affect critically the movement of water
\mbox{molecules.}\looseness=2

\vskip2mm {\it Authors are pleased to thank all the colleagues, with
whom they discussed the problem of dimerization in water. We are
especially thankful to Professor Leonid Bulavin and Professor
Valeriy Pogorelov for numerous discussions. We thank Professor
Georgiy Malenkov for the very fruitful discussion of the results
obtained and the reference to the work of Professor Paul Barnes.  We
are grateful to Professor Paul Barnes for sending us the results
concerning the nature of the interaction between water molecules. We
also thank Professor Mykola Lebovka and Professor Longin Lysetskiy
for the support of our results. We cordially thank all the
participants of the XX International school-seminar ``Spectroscopy
of molecules and crystals'' that took place in 2011 in Beregove. We
owe a lot to the departed Professor Galyna Puchkovska who always
supported us in the investigation of the role and properties of
hydrogen bonds. This work was partially supported by the grant of
SFFR $\sharp$0112U001739.\looseness=-1}

%\vspace*{-2mm}

\vspace*{-5mm}
\rezume{%
П.В. Махлайчук, М.П. Маломуж, І.В. Жиганюк}{ДИМЕРІЗАЦІЯ МОЛЕКУЛ
ВОДИ.\\ МОДЕЛЮВАННЯ МІЖМОЛЕКУЛЯРНОЇ ВЗАЄМОДІЇ\\ НА ОСНОВІ
МУЛЬТИПОЛЬНОГО ПОТЕНЦІАЛУ} {Робота присвячується детальному аналізу
властивостей димерів води. Всі питання, пов'язані з цією проблемою,
досліджуються на основі мультипольного потенціалу, подовженого до
квадруполь-квадрупольного та диполь-октупольного внесків. Отримано
значення всіх основних рівноважних параметрів димеру води: його
геометричних характеристик, енергії основного стану, дипольного і
квадрупольного моментів, частот коливальних станів тощо. Детально
обговорюється адекватність модельних потенціалів міжмолекулярної
взаємодії у воді. Пропонується новий погляд на природу водневого
зв'язку. Результати дослідження ретельно зіставлено з даними
спектроскопічних досліджень і комп'ютерних симуляцій.}

\end{document}